\documentclass[prl,a4paper,twocolumn,showpacs,floatfix,superscriptaddress,amsmath,amsfonts,amssymb,preprintnumbers,citeautoscript,aps]{revtex4-1}

\usepackage{hyperref}
\hypersetup{backref=true,pagebackref=true,hyperindex=true,colorlinks=true,breaklinks=true,
urlcolor=blue,linkcolor=blue,bookmarks=true,bookmarksopen=true,citecolor=red}

\usepackage[T1]{fontenc}\usepackage[latin1]{inputenc}

\usepackage{amsmath}
\usepackage{amssymb}
\usepackage{bbold} % for identity matrix
\usepackage{graphicx}% Include figure files
\usepackage{dcolumn,graphicx,color,booktabs,microtype,afterpage}

\usepackage{bm}% bold math
\usepackage{braket}

\usepackage{times}

\usepackage{color}
\usepackage[normalem]{ulem}

\begin{document}

\title{Orbital Nernst Effect of Magnons}

\author{Li-chuan~Zhang}
\affiliation{Peter Gr\"unberg Institut and Institute for Advanced Simulation, Forschungszentrum J\"ulich and JARA, 52425 J\"ulich, Germany}
\affiliation{Department of Physics, RWTH Aachen University, 52056 Aachen, Germany}
\author{Fabian~R.~Lux}
\affiliation{Peter Gr\"unberg Institut and Institute for Advanced Simulation, Forschungszentrum J\"ulich and JARA, 52425 J\"ulich, Germany}
\affiliation{Department of Physics, RWTH Aachen University, 52056 Aachen, Germany}
\author{Jan-Philipp~Hanke}
\affiliation{Peter Gr\"unberg Institut and Institute for Advanced Simulation, Forschungszentrum J\"ulich and JARA, 52425 J\"ulich, Germany}
\author{Patrick~M.~Buhl}
\affiliation{Institute of Physics, Johannes Gutenberg University Mainz, 55099 Mainz, Germany}
\author{Sergii~Grytsiuk}
\affiliation{Peter Gr\"unberg Institut and Institute for Advanced Simulation, Forschungszentrum J\"ulich and JARA, 52425 J\"ulich, Germany}
\author{Stefan~Bl\"ugel}
\affiliation{Peter Gr\"unberg Institut and Institute for Advanced Simulation, Forschungszentrum J\"ulich and JARA, 52425 J\"ulich, Germany}
\author{Yuriy~Mokrousov}\email[Corresponding author:~]{y.mokrousov@fz-juelich.de}
\affiliation{Peter Gr\"unberg Institut and Institute for Advanced Simulation, Forschungszentrum J\"ulich and JARA, 52425 J\"ulich, Germany}
\affiliation{Institute of Physics, Johannes Gutenberg University Mainz, 55099 Mainz, Germany}

\begin{abstract}
In the past, magnons have been shown to mediate thermal transport of spin in various systems.
Here, we reveal that the fundamental coupling of scalar spin chirality, inherent to magnons, to the electronic degrees of freedom in the system can result in the generation of sizeable orbital magnetization and thermal transport of orbital angular momentum. We demonstrate the emergence of the latter phenomenon of  orbital Nernst effect by referring to the spin-wave Hamiltonian of kagome ferromagnets, predicting that in a wide range of systems the transverse current of orbital angular momentum carried by magnons in response to an applied temperature gradient can overshadow the accompanying spin current. 
We suggest that the discovered effect fundamentally correlates with the topological Hall effect of fluctuating magnets, and it can be utilized in magnonic devices for generating magnonic orbital torques.
\end{abstract}

\maketitle

Spin-heat conversion is a guiding motive in spin caloritronics, which sets out to explore physical phenomena beyond the limits of conventional electronics for energy-efficient information processing~\cite{vzutic2004spintronics,fradkin2010nematic,bauer2012spin,boona2014spin,chumak2015magnon}. In this light, spin-wave excitations in insulating magnetic materials, known as magnons, offer bright prospects as they mediate thermal spin transport via analogs of Seebeck~\cite{geballe1955seebeck,uchida2008observation,xiao2010theory,adachi2011linear} and Nernst effects~\cite{kikkawa2013longitudinal,miyasato2007crossover,kovalev2016spin,zyuzin2016magnon}. While converting temperature gradients into transverse spin currents as a consequence of the Dzyaloshinskii-Moriya interaction (DMI)~\cite{sergienko2006role,heide2008dzyaloshinskii,kovalev2016spin,zyuzin2016magnon}, the spin Nernst effect of magnons is rather inefficient in light materials as it is  proportional to the spin-orbit coupling strength~\cite{cheng2016spin,kovalev2016spin,zyuzin2016magnon}. Recently, it has been suggested that the complex spin arrangement in non-collinear but coplanar magnets provides an alternative route for triggering spin-heat conversion through magnons~\cite{menzel2012information}, which also relies on a DMI-like coupling to the vector spin chirality $\mathbf{S}_i \times \mathbf{S}_j$ among spins $\mathbf{S}_i$ and $\mathbf{S}_j$.

Nowadays, various magnetic phenomena in chiral spin systems are commonly interpreted based on a second flavor of chirality $-$ the scalar spin chirality (SSC)  $\chi_{ijk}=\mathbf S_i \cdot(\mathbf S_j \times\mathbf S_k )$ between local magnetic moments. This type of chirality,  which is inherent to skyrmions~\cite{nagaosa2013topological,seki2012observation,dos2016chirality,lux2018engineering,redies2019distinct} and frustrated magnets~\cite{taguchi2001spin,fujimoto2009hall,diep2013frustrated,owerre2017topological}, has been crucial for understanding the phenomenon of topological Hall effect~\cite{neubauer2009topological,kanazawa2011large,bruno2004topological}. 
Importantly, it was also discovered that the underlying scalar chirality is able to imprint directly on the electronic system by triggering the formation of orbital currents of special kind~\cite{tatara2002chirality,tatara2003permanent}. 
The special property of such {\it topological} orbital currents is their direct sensitivity to the details of spin arrangement and their manifestly non-relativistic origin~\cite{shindou2001orbital,hoffmann2015topological,hanke2016role}. The topological orbital currents have been shown to be pivotal in giving rise to novel exchange interactions in spin systems~\cite{grytsiuk2019topological}, and  the emergence of the corresponding topological orbital moment (TOM) has been demonstrated for various materials~\cite{hanke2017prototypical,lux2018engineering}. In terms of TOM, which is proportional to the SSC with the constant of proportionality known as the topological orbital susceptibility~\cite{hanke2017prototypical,lux2018engineering,grytsiuk2019topological}, the so-called ring exchange~\cite{sen1995large,gritsev2004phase,katsura2010theory} among spins can be naturally interpreted as the Zeeman coupling of TOM to an external magnetic field.   

\begin{figure*}
\centering
\includegraphics[scale=0.33,trim=0 60 10 10]{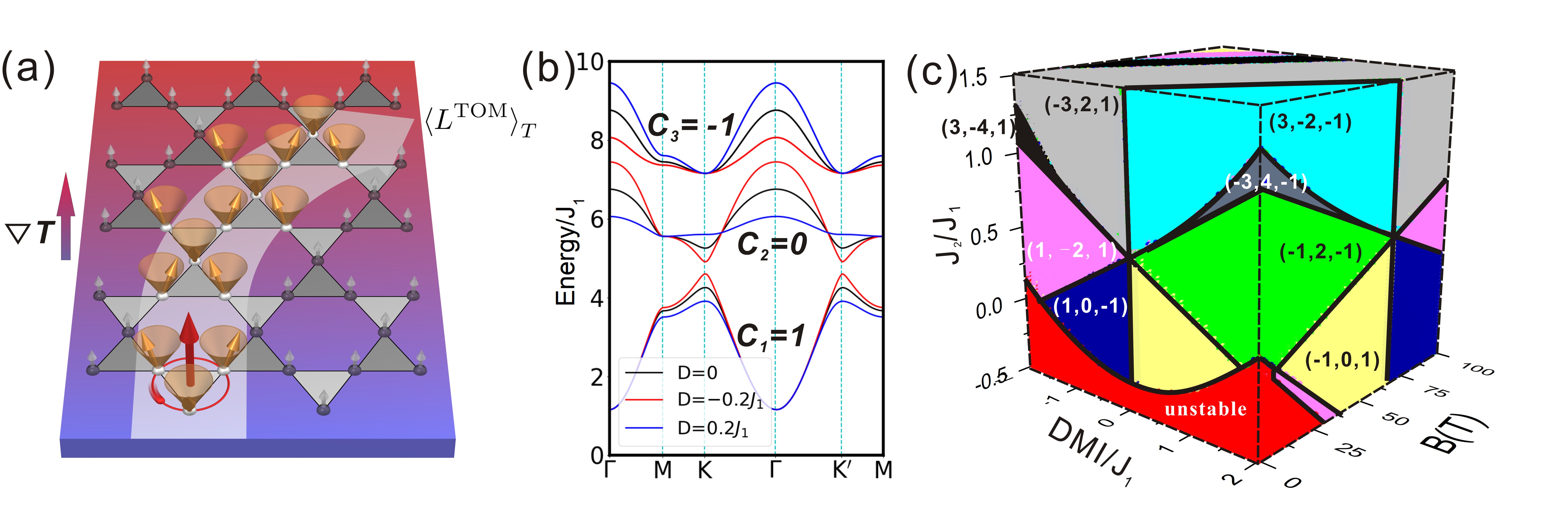}
\caption{(a)~Sketch of the  orbital Nernst effect (ONE) of magnons in a ferromagnet on an example  kagome lattice, where the transport of orbital angular momentum denoted by $\braket{ L^\mathrm{TOM}}$ arising due to deflection of magnons in an applied temperature gradient $\nabla T$ is illustrated with the white arrow. The red arrow indicates the direction of  TOM originating from the local scalar spin chirality. (b)~Magnon dispersion in a magnetic field of 10\,T, where the Chern number of each branch is indicated.  The colors black, red, and blue of the branches stand for the different DMI magnitudes of $0$, $-0.2J_1$, and $0.2J_1$, respectively. (c)~Topological phase diagram of the model as a function of $J_2$, DMI, and external magnetic field. Colors highlight different phases that are characterized by sets $(C_1,C_2,C_3)$ of Chern numbers. The unstable ferromagnetic phase is shown in red. }
\label{fig1}
\end{figure*}

While in the ground state of collinear magnetic systems  the SSC is vanishing, a fundamental question is whether magnonic excitations in such systems can promote it. If yes, the harvesting of SSC by magnons would provide a unique mechanism of coupling directly to the orbital motion of electrons via the effect of topological orbital magnetism. Further, since an applied temperature gradient can drive transverse magnonic scattering, it is
reasonable to ask whether this can result in a magnon ``drag'' of orbital angular momentum. If present, such an effect $-$ coined as the  orbital Nernst effect (ONE) of magnons, see Fig.~\ref{fig1}(a) $-$ would give an ability of driving {\it orbital currents} in addition to currents of spin. This lays out a conceptually new way of generating currents of angular momentum in magnonic systems which could be utilized, e.g., for exerting magnonic orbital torques~\cite{go2018intrinsic,go2019orbital}.

Here, we address these questions by referring to an effective Hamiltonian of spin waves of a ferromagnet on a two-dimensional kagome lattice as given by
\begin{equation}
\begin{split}
    H=&-\frac{1}{2}\sum_{ij}J_{ij}  \mathbf{S}_i \cdot \mathbf{S}_j -\frac{1}{2}\sum_{ij}\mathbf{D}_{ij} \cdot (\mathbf{S}_i\times \mathbf{S}_j)\\  &-\mathbf{B}\cdot \kappa^\mathrm{TO}\sum_{ijk} \hat{\mathbf e}_{ijk} [\mathbf S_i \cdot (\mathbf S_j \times \mathbf S_k)] - \mathbf{B}\cdot\sum_i \mathbf{S}_i  \, ,
\end{split}
\label{Ham}
\end{equation}
 where $J_{ij}$  mediates the Heisenberg exchange between spins $\mathbf{S}_{i}$ and $\mathbf{S}_{j}$ on sites $i$ and $j$, the second term is the antisymmetric DMI quantified by vectors $\mathbf{D}_{ij}$, and the fourth term couples the spins to an external magnetic field $\mathbf B$. In addition, we extend the Hamiltonian by the ring-exchange term in Eq.~\eqref{Ham} to include explicitly the interaction between the magnetic field and the TOM. This term is given by the product of the SSC $\chi_{ijk}$ 
 and the topological orbital susceptibility $\kappa^\mathrm{TO}$, which accumulates the effect of the underlying electronic structure. For a given triangle of spins, the direction of TOM is along the normal $\hat{\mathbf e}_{ijk}$ of the oriented triangle~\cite{hanke2017prototypical,grytsiuk2019topological}. Owing to the symmetry of the planar kagome lattice, the TOM and the DMI vectors are perpendicular to the film plane (along the $z$-direction), along which we also apply the external magnetic field of magnitude $B$. 
We consider in our analysis only nearest-neighbor interactions except for the Heisenberg term, where we include next-nearest neighbors as well. We set the nearest-neighbor Heisenberg coupling to $J_1=1$\,meV, the next-nearest neighbor strength amounts to $J_2=0.1\,J_1$ unless stated otherwise, the spin-moment length $S$ is fixed to $1\,\mu_\mathrm{B}$, and the topological orbital susceptibility $\kappa^\mathrm{TO}$ is chosen as $-0.5\,\mu_\mathrm{B}^{-2}$ $-$ a value motivated by recent material studies~\cite{hanke2016role,grytsiuk2019topological}. 

 Linear spin-wave theory~\cite{toth2015linear,li2017dirac,dos2018spin,mook2016tunable,mook2014magnon} is used to obtain the eigenvalues and eigenvectors of the above Hamiltonian, which we reformulate first in terms of bosonic ladder operators $a_i$ and $a_i^\dagger$ via the Holstein-Primakoff transformation~\cite{holstein1940field}. In the resulting spin-wave Hamiltonian, we keep only terms that are quadratic in the ladder operators, thus neglecting higher-order spin excitations. Then, the SSC $\chi_{ijk}$, coupling directly to the magnetic field in Eq.~\eqref{Ham}, can be expressed as
\begin{equation}
    \chi_{ijk} =  iS^2\,(a_i^{\dag}a_j-a_ia_j^{\dag}+a_j^{\dag}a_k-a_ja_k^{\dag}+a_k^{\dag}a_i-a_ka_i^{\dag}) \, .
    \label{eq:2}
\end{equation}
To map from real to momentum space, we perform a Fourier transform of the bosonic ladder operators, which leads to the Hamiltonian matrix $H(\mathbf k)$ at the spin-wave vector $\mathbf k=(k_x,k_y)$,
  which is diagonalized to obtain the eigenvectors and the energy spectrum of the spin waves. We address the topological character of the magnonic bands by computing the Chern number $C_n$, given by 
$C_n=\frac{1}{2\pi}\int\Omega_{n\mathbf k}^{xy} \,d\mathbf k$, where the integral is performed over the Brillouin zone (BZ), and $\Omega_{n\mathbf k}^{xy}$ represents 
the magnon Berry curvature of the $n$th spin-wave branch, given by
\begin{equation}
 \Omega_{n\mathbf k}^{xy}=-2\,\mathrm{Im}\sum_{m\neq n}
 \frac{\braket{\Psi_{n\mathbf{k}}|\frac{\partial H(\mathbf{k})}{\partial k_x}|\Psi_{m\mathbf{k}}}\braket{\Psi_{m\mathbf{k}}|\frac{\partial H(\mathbf{k})}{\partial k_y}|\Psi_{n\mathbf{k}}}}{(\epsilon_{n\mathbf{k}}-\epsilon_{m\mathbf{k}})^2} \, ,
 \label{curvature}
 \end{equation}
where $|\Psi_{n\mathbf k}\rangle$ is an eigenstate of the spin-wave Hamiltonian with the energy $\epsilon_{n\mathbf k}$~\cite{SupplementalMaterial}.

\begin{figure*}
\centering
\includegraphics[scale=0.32,trim=10 20 0 0]{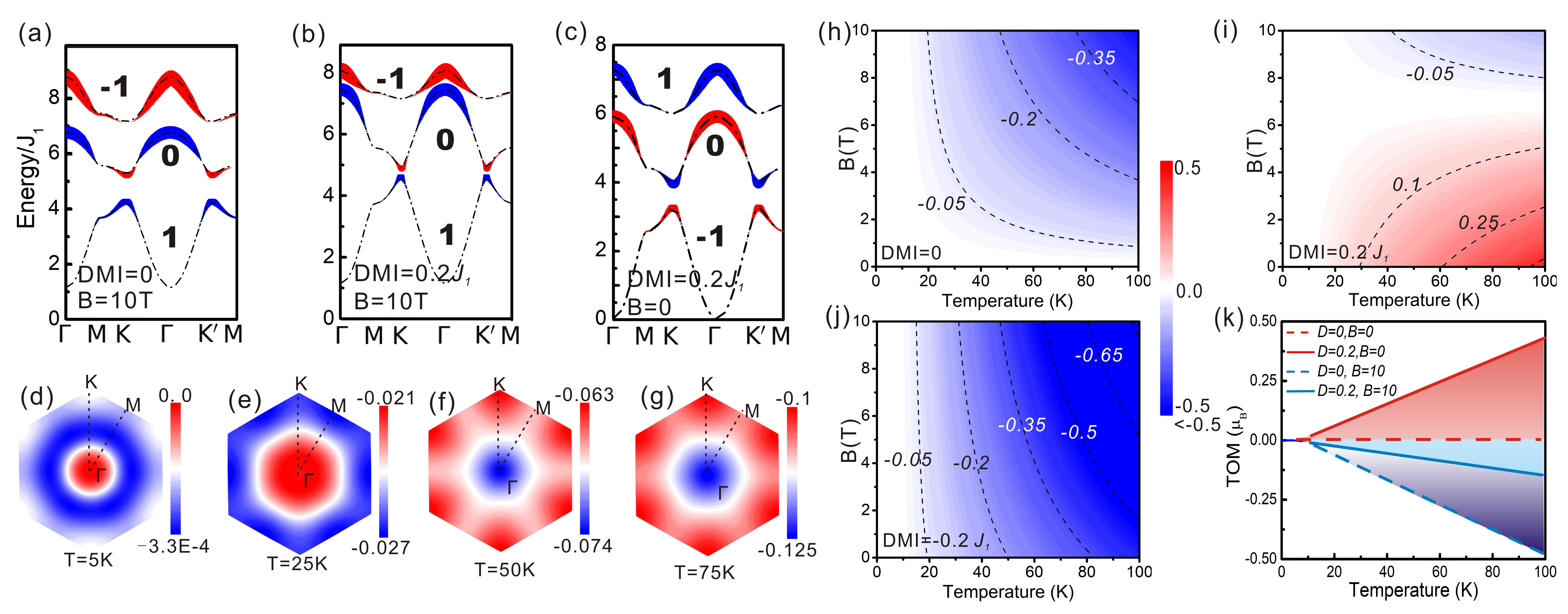}
\caption{(a--c)~Fat band analysis for the ferromagnetic kagome lattice. Red and blue colors represent positive and negative sign of the local TOM $L_{n\mathbf k}^\mathrm{TOM}$, respectively, and the line thickness denotes the corresponding magnitude. Bold integers indicate the Chern numbers of the spin-wave bands. (d--g)~Distribution of the local TOM in the Brillouin zone for different temperatures, after summing over all magnon branches weighted by the Bose distribution. The color map is in units of $\mu_\mathrm{B}$, and the model parameters of panel~(a) are used. (h--k)~Overall TOM of the spin-wave system as a function of magnetic field and temperature. The panels (h--j) present phase diagrams for the DMI strengths of $0$, $0.2J_1$, and $-0.2J_1$, respectively, with the color map indicating the net TOM in units of $\mu_\mathrm{B}$ per unit cell. In (k),  solid and dotted lines correspond to DMI strengths of $0$ and $0.2J_1$, respectively, and the magnetic field is given in tesla.}
\label{fig2}
\end{figure*}

We start with an analysis of the dispersion of the three spin-wave branches of our model in the presence of an external magnetic field of $10$\,T, presented in Fig.~\ref{fig1}(b). In the absence of DMI, the different magnon bands exhibit Chern numbers $1$, $0$, and $-1$ solely due to the coupling of the magnetic field to the SSC manifesting in a non-zero TOM carried by the magnons, as we show below.  
By including the effect of DMI,  Fig.~\ref{fig1}(b), we find that the coupling to the vector spin chirality modifies the dispersion without changing the topology of the bands for this choice of parameters. While the microscopic origin of interactions with vector and scalar spin chiralities which enter Eq.~\eqref{Ham} is fundamentally different, their role for the resulting magnon dispersion is rather similar at the level of linear spin-wave theory.
Based on the obtained spin-wave spectra and Berry curvature calculations, we present in Fig.~\ref{fig1}(c) the complete topological phase diagram as a function of the model parameters entering in the  Hamiltonian. We distinguish the phases using the set of Chern numbers $(C_1,C_2,C_3)$ of the  magnon branches. Sampling the nearest-neighbor coupling $J_2$, the DMI strength, and the magnitude of the $B$-field, 
we identify eight non-trivial phases in addition to an unstable ferromagnetic state. These phases come in pairs with an opposite overall sign in the set of Chern numbers.

To uncover the role of magnons in giving rise to orbital magnetism of the electrons via the mechanism of scalar spin chirality outlined above, we evaluate the local TOM of the $n$th magnon branch according to $L_{n\mathbf k}^\mathrm{TOM}=\kappa^\mathrm{TO} \braket{\Psi_{n\mathbf k}|\chi(\mathbf k)|\Psi_{n\mathbf k}}$. Figure~\ref{fig2}(a--c) illustrates the value of the local TOM of the magnon branches as represented by the line thickness. While either finite DMI or $B$-field are necessary to activate the local TOM, the $\Gamma$ point typically hosts the minima and maxima of $L_{n\mathbf k}^\mathrm{TOM}$ in our model. Specifically, the local TOM of the lowest spin-wave branch reaches its global minimum at $\Gamma$ whereas the higher magnon bands carry the maximal values as they correspond to precessional modes with an innately larger  SSC. Following the evolution of the Chern numbers in Fig.~\ref{fig2}(a--c), we show that the complex interplay between DMI and the orbital Zeeman coupling modifies not only the magnon topology but imprints also on the local TOM. In particular, the ordering of the states with positive and negative sign of $L_{n\mathbf k}^\mathrm{TOM}$ is inverted during the topological phase transition.

We turn now to the effect of thermally excited magnons on TOM. In Fig.~\ref{fig2}(d--g) we analyze the sum of the local emergent orbital moment weighted by the occupation number of each spin-wave branch at a given temperature, i.e., $\ell(\mathbf k)=\sum_n L_{n\mathbf k}^\mathrm{TOM} n_\mathrm{B}(\epsilon_{n\mathbf k})$. Here, the magnons follow the Bose distribution function $n_\mathrm{B}(\epsilon)=[\exp{(\beta\epsilon)}-1]^{-1}$ with $\beta=1/k_\mathrm{B}T$. Depending on the temperature $T$, the number of excited magnons is different in each branch, which leads to a non-trivial distribution of the microscopic quantity $\ell(\mathbf k)$ in momentum space, as shown in Fig.~\ref{fig2}(d--g) for the model with finite $B$-field but zero DMI. 
In the low-temperature regime (e.g., for $T=5$\,K), only the $\Gamma$-point magnons from the first branch can be excited, leading only to small local contributions around the BZ center. As the temperature is increased, for example, to $T=25$\,K, all spin-wave states from the first branch are excited such that $\ell(\mathbf k)$ peaks in the $M$ point with moderate magnitude as shown in Fig.~\ref{fig2}(e). If additionally magnons from the higher branches contribute at elevated temperatures, the maximum of $\ell(\mathbf k)$ occurs at the $\Gamma$ point, where the local TOM of the corresponding magnon states is the largest.

The overall TOM of the spin-wave system at given $T$ can be then obtained as an integral over the Brillouin zone:
\begin{equation}
\begin{split}
&\braket{L^\mathrm{TOM}}_T = \int \ell(\mathbf k) \, d\mathbf k = \sum_n \int n_\mathrm{B}(\epsilon_{n\mathbf k}) \, L_{n\mathbf k}^\mathrm{TOM}\, d\mathbf k,
\end{split}
\label{TOM}
\end{equation} 
where $\braket{L^\mathrm{TOM}}_T$ is the total emergent orbital moment carried by thermally activated magnons per unit cell~\cite{SupplementalMaterial}.
Figure~\ref{fig2}(h--j) illustrates the dependence of the overall TOM on magnetic field and temperature for various DMI coupling strengths. On the one hand, as more magnons become available to carry the TOM, higher temperatures enhance the magnitude of $\braket{L^\mathrm{TOM}}_T$ in the spin-wave system. On the other hand, the roles of orbital Zeeman coupling and DMI are intertwined in generating TOM. 
For example, while TOM locally vanishes at zero DMI and $B$-field, a DMI with positive coupling strength generally counteracts the effect of the magnetic field on TOM if $\kappa^{\rm TO}$ is negative. 
For non-trivial choices of these parameters, however, Fig.~\ref{fig2}(k) illustrates that at low $T$ the total TOM increases linearly, and, depending on the value of  $\kappa^{\rm TO}$, it can reach a sizeable magnitude. The sign of $\braket{L^\mathrm{TOM}}_T$ correlates with the ordering of the topological spin-wave bands and their respective Chern numbers.

Answering the first question posed in the beginning, our analysis demonstrates that a finite TOM, stemming from orbital electronic currents in the kagome ferromagnet, can be triggered by thermally activated magnons. This observation suggests that TOM is intimately linked to thermal spin transport which is mediated by the coupling of the SSC to the bath of electrons in these systems. As a consequence, the well-known magnon Nernst effect acquires a novel and fundamentally distinct contribution that we coin the orbital Nernst effect (ONE) of magnons, which is illustrated in Fig.~\ref{fig1}a. The phenomenon of ONE relates spatial temperature gradients to the emergence of topological orbital currents via $j_x^\mathrm{TOM}=\kappa^{xy}_\mathrm{ONE}(\nabla T)_y$, where $\kappa_\mathrm{ONE}^{xy}$ stands for the topological orbital Nernst conductivity, which within the semiclassical theory reads
\begin{equation}
\begin{split}
    \kappa_\mathrm{ONE}^{xy}=-\frac{k_\mathrm{B}}{(2\pi)^2\mu_\mathrm{B} }\sum_n\int c_1(n_\mathrm{B}(\epsilon_{n\mathbf k}))\,\Omega_{n\mathbf k}^{xy}
    L_{n\mathbf k}^\mathrm{TOM} \, d\mathbf k,
\end{split}
\label{eq:tone}
\end{equation}
where 
$c_1(\tau)=\int_0^{\tau}\ln[(1+t)/t]dt=(1+\tau)\ln(1+\tau)-\tau\ln \tau$. In contrast to the usual spin Nernst effect of magnons~\cite{matsumoto2011theoretical,murakami2016thermal}, the conductivity in Eq.~\eqref{eq:tone} characterizing the ONE depends explicitly on the local TOM of the magnon branches.

\begin{figure}
\centering
\includegraphics[scale=0.23,trim=0 10 0 10]{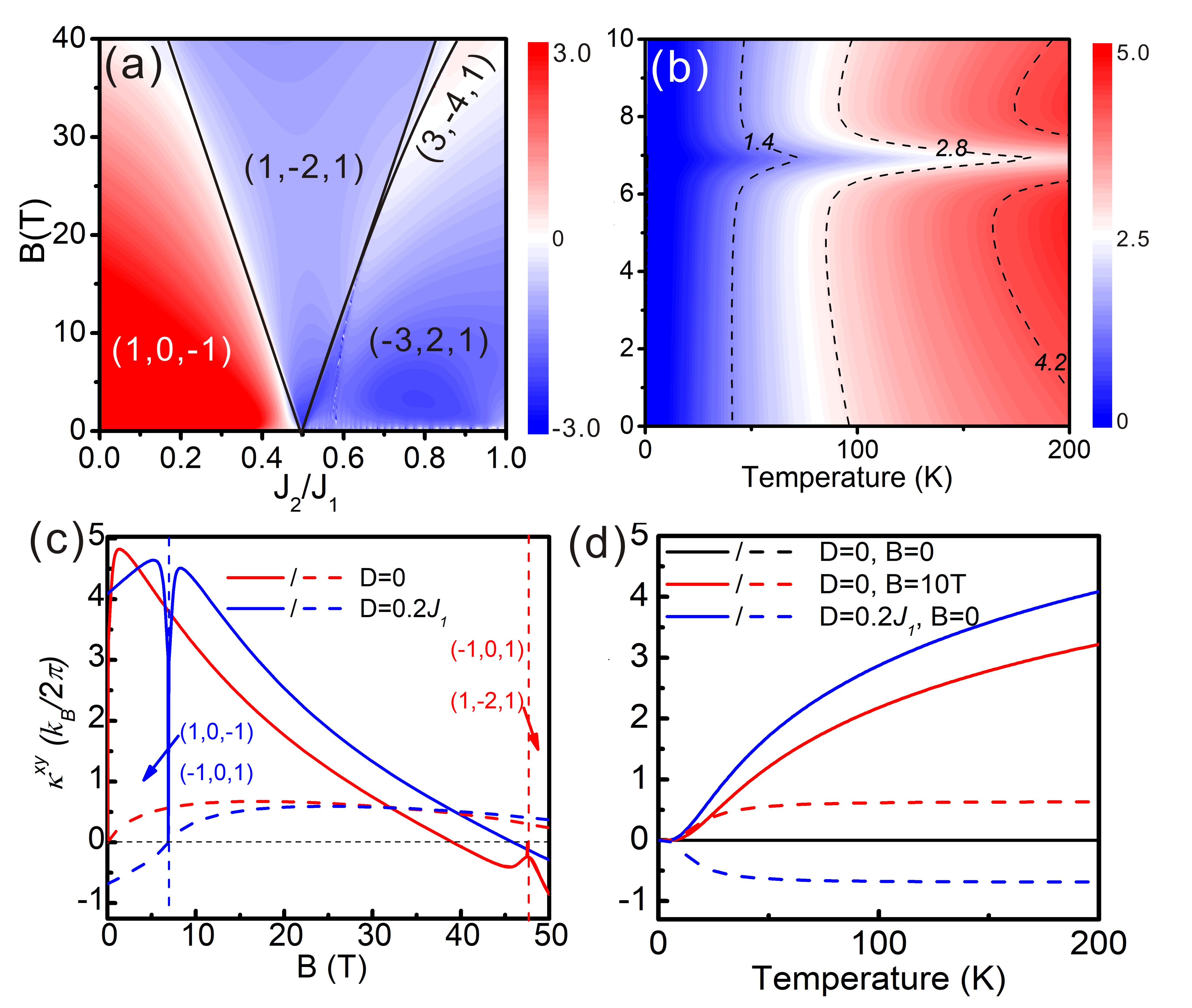}
\caption{ (a)~Phase diagram of ONE. Dependence of the orbital Nernst conductivity $\kappa_\mathrm{ONE}^{xy}$ on $B$ and $J_2$ at $T=200$\,K and zero DMI strength. Solid black lines are the boundaries between different topological phases characterized by the Chern numbers of the three magnon branches.  (b)~Orbital Nernst conductivity as a function of $B$ and temperature $T$ for the model with DMI strength of  $0.2J_1$. (c,d)~Comparison of the $\kappa_\mathrm{ONE}^{xy}$ (solid lines) and magnon Nernst conductivity $\kappa_\mathrm{N}^{xy}$ (dashed lines). (c)~The  $\kappa_\mathrm{ONE}^{xy}$ and $\kappa_\mathrm{N}^{xy}$ as a function of $B$ for the model at 200~K with DMI strength of 0 (red) and $0.2J_1$ (blue). The different topological phases are distinguished with a thin vertical line. (d)~Parameters $\kappa_\mathrm{ONE}^{xy}$ and $\kappa_\mathrm{N}^{xy}$ as a function of $T$ for different strengths of the DMI and $B$.}
\label{fig3}
\end{figure}

Answering the second posed fundamental question, below we reveal the existence of this effect. In Fig.~\ref{fig3} we summarize the non-trivial dependence of the ONE on temperature and on the model parameters, as well as its correlation with the topology of the magnon bands. Although the ONE has a distinct microscopic origin in the orbital electron-magnon coupling, our prediction is that the corresponding conductivity can reach the order of $k_\mathrm{B}/\pi$~\footnote{If we assume a distance of $5$~\AA{} between two kagome layers, an orbital Nernst conductivity of $k_\mathrm{B}/(2\pi)$ is equivalent to the value $4.394\times 10^{-15}$~Jm$^{-1}$K$^{-1}$, or $66786$~$(\hbar/e)\mu$Acm$^{-1}$K$^{-1}$.}, which is comparable to the values known for the spin Nernst effect of magnons or spin Nernst effect of electrons~\cite{kovalev2016spin,zyuzin2016magnon,cheng2016spin,mook2014magnon,mook2019thermal,geranton2015spin,long2016giant}. This underlines the strong potential of ONE for the realm of spincaloritronics. 

We show in Fig.~\ref{fig3}(a) and Supplementary Figure~S6(b) that both DMI and the coupling of external magnetic field to the SSC can generate a finite Nernst conductivity. Comparing the two panels in more detail, we note that the sign of $\kappa_\mathrm{ONE}^{xy}$ is the same in topological phases for which the sets of Chern numbers differ by a global sign. This invariance stems from the product of the two microscopic quantities in Eq.~\eqref{eq:tone}, each of which changes its individual sign as the Chern numbers of the spin-wave branches are reversed. Still, as exemplified in Fig.~\ref{fig3}(a,c), the ONE is characteristic to the non-trivial magnon topology of distinct phases.  
Close to topological phase transitions, the ONE changes abruptly and thus behaves rather differently compared to thermal Hall and magnon Nernst effects, see Fig.~\ref{fig3}(c). As a consequence, the conductivity $\kappa_\mathrm{ONE}^{xy}$ can in principle reach very large values near the phase boundary. 
Since ONE is absent without the $B$-field and DMI (see Fig.~S9(a)), the peak structure in Fig.~\ref{fig3}(b,c) for a magnetic field of about $7$\,T can be understood as a result of the competition between the effects of orbital Zeeman coupling and DMI, 
which results in a strongly suppressed ONE.
On the other hand, Fig.~\ref{fig3}(c,d) and Fig.~S9(a) reveal the qualitative difference in the temperature dependence of ONE and conventional Nernst effect.

The mechanism of magnon-driven chirality accumulation uncovered in this work has far-reaching consequences for the transport properties of systems which exhibit such chirality. For example, it will result
in the generation of topological Hall or topological spin Hall effect of the underlying electronic bath~\cite{bruno2004topological,buhl2017topological}, which can be easily
estimated based on the knowledge of the so-called topological Hall constant of the system~\cite{franz2014real,buhl2017topological}, and which will contribute to the temperature dependence of the anomalous Hall conductivity even  in nominally collinear magnets~\cite{ishizuka2018spin}.
On the other hand, the corresponding emergent orbital magnetism associated with non-vanishing net chirality of magnons brings the orbital angular momentum variable into the game of magnon-based spincaloritronics, which is conventionally associated with generation and transport of spin. Unleashing the orbital channel for the magnon-mediated effects poses a key question of the role of orbital magnetism for the temperature-dependent magnetization dynamics, however, it also opens a number of exciting possibilities for direct applications. For example, in analogy to the spin-orbit torques~\cite{liu2012current,liu2012spin,miron2010current,garello2013symmetry}, we envisage that the flow of orbital angular momentum generated by magnons can be used to generate sizeable orbital accumulation and orbital torques on adjacent magnets. On the other hand, given the apparent sensitivity of the effects considered here to the topology of magnonic bands, we suggest that accessing the magnon-mediated dynamics of orbital properties can serve as a unique tool of tracking the topological dynamics of magnons. 

\begin{acknowledgments}
We acknowledge fruitful discussions with Marjana Le\v{z}ai\'c, Olena Gomonay and Dongwook Go. L.-C.~Zhang acknowledges support from China Scholarship Council (CSC) (No. [2016]3100). We  acknowledge  funding  under SPP 2137 ``Skyrmionics" (project  MO  1731/7-1)  of  Deutsche  Forschungsgemeinschaft (DFG). We also gratefully acknowledge the J\"ulich Supercomputing Centre and RWTH Aachen University for providing computational resources under project jiff40.
\end{acknowledgments}

\bibliographystyle{apsrev4-1}
\bibliography{TOM}\vspace{-3pt}
\end{document}